\documentclass{mn2e}

\usepackage{amsfonts}
\usepackage{amsmath}
\usepackage{graphicx}

\title[Satellite luminosities in galaxy groups]
{Satellite luminosities in galaxy groups}

\author[R. A. Skibba, R. K. Sheth \& M. Martino]
{Ramin A. Skibba$^{1,2}$\thanks{E-mail:  skibba@mpia.de (RAS); shethrk@physics.upenn.edu (RKS)}
 Ravi K. Sheth$^{3}$ \& Matthew C. Martino$^3$\footnotemark[1]\\
$^{1}$Department of Physics \& Astronomy, University of Pittsburgh, 
      Pittsburgh, PA 15260, USA\\
$^{2}$Max Planck Institute for Astronomy, K\"{o}nigstuhl 17,
      D-69117 Heidelberg, Germany\\
$^{3}$Department of Physics \& Astronomy, University of Pennsylvania, 
      Philadelphia, PA 19104, USA}

\newcounter{appfig}

\begin{document}
\pagerange{\pageref{firstpage}--\pageref{lastpage}}

\maketitle

\label{firstpage}

\begin{abstract}
Halo model interpretations of the luminosity dependence of galaxy 
clustering assume that there is a central galaxy in every 
sufficiently massive halo, and that this central galaxy is very 
different from all the others in the halo.  The halo model 
decomposition makes the remarkable prediction that the mean 
luminosity of the non-central galaxies in a halo should be almost 
independent of halo mass---the predicted increase is about twenty 
percent while the halo mass increases by a factor of more than twenty.  
In contrast, the luminosity of the central object is predicted to 
increase approximately linearly with halo mass at low to 
intermediate masses, and logarithmically at high masses.  
We show that this weak, almost non-existent mass-dependence of the 
satellites is in excellent agreement with the satellite population 
in group catalogs constructed by two different collaborations.
This is remarkable, because the halo model prediction was made 
without ever identifying groups and clusters.  The halo model also 
predicts that the number of satellites in a halo is drawn from a 
Poisson distribution with mean which depends on halo mass.  This, 
combined with the weak dependence of satellite luminosity on halo 
mass, suggests that the Scott effect,
such that the luminosities of very bright galaxies are merely the  
statistically extreme values of a general luminosity distribution, 
may better apply to the most 
luminous {\em satellite} galaxy in a halo than to BCGs.  
If galaxies are identified with dark halo substructure at the 
present time, then central galaxies should be about 4 times more 
massive than satellite galaxies of the same luminosity, whereas the 
differences between the stellar mass-to-light ratios should be smaller.  
Therefore, a comparison of the weak lensing signal from central 
and satellite galaxies of the same luminosity should provide 
useful constraints on these models.  We also show how the halo 
model may be used to constrain the stellar mass associated with 
intracluster light:  the mass fraction in the ICL is expected to 
increase with increasing halo mass.  
\end{abstract}

\begin{keywords}
methods: analytical - galaxies: formation - galaxies: haloes -
dark matter - large scale structure of the universe 
\end{keywords}

\section{Introduction}
The halo model (see Cooray \& Sheth 2002 for a review) has become 
the preferred language in which to interpret measurements of galaxy 
clustering.  Recently, Zehavi et al. (2005) have expressed the 
luminosity dependence of clustering in the Sloan Digital Sky Survey
(SDSS, York et al. 2000) Second Data Release (DR2, Abazajian et al. 2004)
in terms of the halo model.
Skibba et al. (2006) show that, if Zehavi et al.'s halo model decomposition is correct, 
then the luminosity of the central galaxy in a halo depends strongly 
on halo mass, whereas the luminosities of satellite galaxies depend 
only weakly on the masses of their host haloes.  The main goal of 
this paper is to test this prediction.  
We do this in Section~\ref{testLsat} by studying the satellite 
population in the group catalog provided by Berlind et al. (2006).  
The abundance of groups decreases and the clustering strength 
increases with increasing richness, as expected (Berlind et al. 2007).  
This suggests that the test we perform is unlikely to have been 
biased by incompleteness effects in the catalog.  As an additional 
check, we show that the satellite population in the group catalogs 
of Yang et al. (2005a) are similar to those from Berlind et al. 

Dark matter haloes have substructure (e.g. Tormen 1997; 
Tormen, Diaferio \& Syer 1998; Gao et al. 2004a).  If we identify 
subhaloes with satellite galaxies (e.g. Kravtsov et al. 2004; 
Conroy et al. 2007), then the halo model makes specific predictions 
about how center and satellite galaxies of the same luminosity differ; 
this difference is the subject of Section~\ref{compareM2L}.  
These predictions can also be tested by studying how stellar 
and total mass-to-light ratios depend on environment; 
how the luminosity function of clusters (after removing 
the BCG) depends on cluster richness; 
and how the amount of intracluster light depends on cluster richness.   
The connections between these tests and the halo model are discussed 
in a final section which summarizes our findings.  
Throughout, we assume a spatially flat cosmology with 
$\Omega_0=0.3$, $\Lambda_0=1-\Omega_0$ and $\sigma_8=0.9$, and 
we write the Hubble constant as $H_0=100h$~km~s$^{-1}$~Mpc$^{-1}$.  

An Appendix presents a few inconsistencies between the halo 
model of Zehavi et al. (2005) and the group catalog of 
Berlind et al. (2006).  It argues that while these may be due 
to Zehavi et al.'s assumption that $\sigma_8=0.9$, they are 
unlikely to invalidate our findings.  

\section{The halo model and group galaxies}\label{testLsat}
The halo model decomposition provides a prescription for how 
the galaxy population in a halo depends on halo mass.  In practice, 
halo mass is not an observable, so comparison of this prediction 
with the objects in a group catalog is not straightforward.  
However, the halo model decomposition can be re-written so that 
observable quantities are predicted:  these include the number 
density of groups containing $N$ galaxies more luminous than 
some threshold luminosity, as well as the average luminosities 
of the central and satellite galaxies as a function of $N$.  
Specifically, 
\begin{equation}
 n_{\rm grp}(N) = \int_{M_{\rm min}(L_{\rm min})}^\infty
  {\rm d}M\, {{\rm d}n(M)\over {\rm d}M}\, p(N|M),
 \label{ngrpN}
\end{equation}
\begin{equation}
 \bigl< L_{\rm cen}|N\bigr> =  \int_{M_{\rm min}(L_{\rm min})}^\infty
    {\rm d}M\, {{\rm d}n(M)\over {\rm d}M}
      {p(N|M)\, L_{\rm cen}(M)\over n_{\rm grp}(N)}
 \label{LcenN}
\end{equation}
and 
\begin{equation}
 \bigl< L_{\rm sat}|N,L_{\rm min}\bigr> = 
	 \int_{M_{\rm min}(L_{\rm min})}^\infty {\rm d}M\, 
              {{\rm d}n(M)\over {\rm d}M}\, 
	    {p(N|M)\, \bigl< L_{\rm sat}|M,L_{\rm min}\bigr>
             \over n_{\rm grp}(N)},
 \label{LsatN}
\end{equation}
where $dn(M)/dM$ is the halo mass function (we use the parametrization 
given by Sheth \& Tormen 1999), 
and the distribution $p(N|M)$ has mean 
\begin{equation}
 \Bigl<N|M,\ge L\Bigr> = 1 + \Bigl<N_{\rm sat}|M,\ge L\Bigr> 
               = 1 + \left[{M\over M_1(L)}\right]^{\alpha(L)},
 \label{sdssNg}
\end{equation}
with $N_{\rm sat}$ drawn from a Poisson distribution (Zehavi et al. 2005).
Here $M_1(L)\approx 23\,M_{\rm min}(L)$, where $M_{\rm min}$ denotes 
the minimum mass required to host a galaxy of luminosity $L$ or 
greater, and $\alpha\approx 1$.  

The minimum mass scales with $r$-band $L$ 
(our $r$-band is actually the SDSS $r$ filter shifted to $z=0.1$, 
sometimes denoted $^{0.1}r$) as 
\begin{equation}
 \left({M_{\rm min}\over 10^{12}h^{-1}M_\odot}\right)\approx 
    \exp\left(\frac{L}{1.1\times10^{10}\,h^{-2}L_\odot}\right) - 1,  
 \label{MLapprox}
\end{equation}
so 
\begin{equation}
 {L_{\rm cen}\over 1.1\times 10^{10}\,h^{-2}L_\odot}\approx 
  \ln\left(1 + {M\over 10^{12}\,h^{-1}M_\odot}\right)
 \label{LMapprox}
\end{equation}
(Skibba et al. 2006), assuming $M_{\odot r}=4.76$ (Blanton et al. 2003).
Note that the luminosity of the central object is predicted to 
increase linearly with halo mass when $M\ll 10^{12}\,h^{-1}M_\odot$, 
but the increase is only logarithmic at larger $M$ (also see 
Tinker et al. 2005).  This is qualitatively consistent with the 
findings of Lin \& Mohr (2004), Yang et al. (2005b) and Cooray (2006).  

The mean satellite luminosity is given by 
\begin{equation}
 \Bigl< L_{\rm sat}|M,L_{\rm min}\Bigr> = L_{\rm min} + 
 \int_{L_{\rm min}}^\infty dL \, {[M/M_1(L)]^{\alpha(L)}\over 
                                  [M/M_1(L_{\rm min})]^{\alpha(L_{\rm min})}}.
 \label{LsatM}
\end{equation}
Figure~2 of Skibba et al. (2006) shows that this is a much weaker 
function of $M$ than is $L_{\rm cen}$.  To see why, notice that if 
$\alpha(L)$ were independent of $L$, then the mean satellite luminosity 
would be independent of $M$.  This suggests that the $L$-dependence of 
$\alpha$ reflects the mass dependence of satellite luminosities:  
the halo model prediction that satellite luminosities depend only 
weakly on halo mass is a consequence of the fact that $\alpha$ is 
only weakly dependent on $L$.  

\begin{figure}
 \centering
 \includegraphics[width=\hsize]{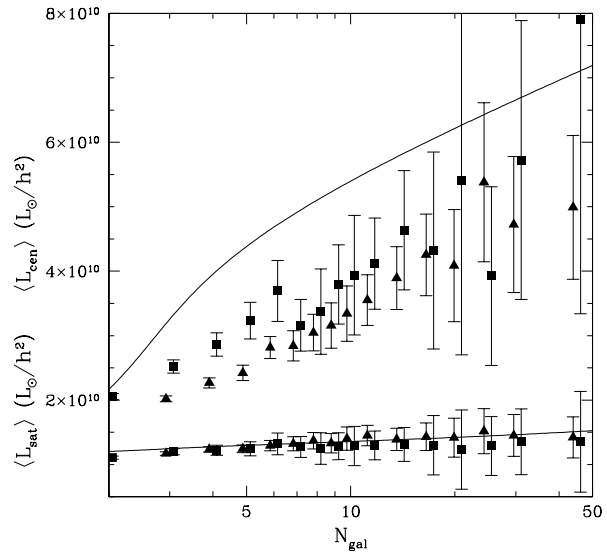} 
 \caption{Comparison of the mean satellite luminosity in groups 
          containing $N_{\rm gal}$ members each more luminous than $M_r\le -19.9$ 
          (lower set of symbols with error bars).  Triangles and squares 
          show results from the group catalogs of Berlind et al. (2006) 
          and Yang et al. (2005a), respectively, and the points have been 
          slightly offset in $\mathrm{log}(N_{\rm gal})$, for clarity.
          Lower solid line shows the halo model prediction of this 
          quantity (equation~\ref{LsatN}).  
          Upper set of symbols with errors show that the mean 
          central luminosity $L_{\rm cen}$ is a stronger function 
          of $N$ than is $L_{\rm sat}$. Solid line shows that the 
          halo model (equation~\ref{LcenN}) correctly predicts the 
          stronger $N$ dependence, but overpredicts the actual 
          luminosities.}
 \label{berlindLsat}
\end{figure}

Figure~\ref{berlindLsat} compares equation~(\ref{LsatN}) with the 
mean satellite luminosity in the $M_r\le -19.9$ group catalog of 
Berlind et al. (2006).  This catalog is drawn from the SDSS Fourth 
Data Release (DR4, Adelman-McCarthy et al. 2006); it is volume-limited over 
the redshift range $0.015<z<0.100$, and consists of 21301 galaxies 
in 4119 groups having three or more members.  The groups were 
identified using a redshift-space friends-of-friends algorithm, 
which was tuned to identify galaxy systems that occupy the same 
host dark matter halo, based on halo occupation distribution models,
the group multiplicity function, and distributions of group sizes and 
velocity dispersions.  See Berlind et al. for details of the 
algorithm and resulting catalogs.  

We also compare to a group catalog of Yang et al. (2005a) and 
Weinmann et al. (2006), which is drawn from the SDSS DR2 
(Abazajian et al. 2004).  
When restricted to the same volume limited cuts, it consists of 
10475 galaxies in 3260 groups having two or more members.  
The groups were identified using a different redshift-space 
friends-of-friends algorithm.  See Yang et al. for details of 
the algorithm, and Weinmann et al. for their application to the SDSS.  
The two sets of symbols in Figure~\ref{berlindLsat} show that, 
whereas the Yang et al. catalog has slightly brighter central 
galaxies at fixed $N$, the two group catalogs predict almost 
the same scaling of $\langle L_{\rm sat}\rangle$ with $N$.  

The halo model parameters associated with this luminosity 
threshold are $M_{\rm min}\approx 10^{12}h^{-1}M_\odot$ and 
$\alpha = 1.13$, according to the halo occupation distribution 
analysis of Zehavi et al. (2005). 
The symbols in Figure~\ref{berlindLsat} show that $L_{\rm sat}$ 
increases only very weakly with $N$ (essentially because it 
increases only weakly with $M$), whereas $L_{\rm cen}$ is a 
stronger function of $N$.  The solid curves show that, although 
equation~(\ref{LcenN}) overpredicts the luminosities of the central 
galaxies, equation~(\ref{LsatN}) reproduces the scaling of 
$L_{\rm sat}$ with $N$ quite well.  
Although the agreement is not perfect, it is nevertheless 
remarkable, because the halo model decomposition into central 
and satellite objects is done without ever identifying groups and 
clusters in the SDSS dataset.  Therefore, this agreement with the 
satellite population in an actual group catalog represents a 
nontrivial success of the approach.  (The discrepancy for the 
central galaxies is not a central issue of this paper---it is 
discussed further in the Appendix.)

Equations~(\ref{sdssNg}) and~(\ref{MLapprox}) imply that 
the satellite galaxy luminosity function is given by 
\begin{equation}
 \phi_{\rm sat}(\ge L|M) =
    \left({M_{12}\over 23}\right)^{\alpha(L)}
 \Bigl[\exp(L_{10}) - 1\Bigr]^{-\alpha(L)}
 \label{cumLsat}
\end{equation}
where $M_{12} = M/10^{12}h^{-1}M_\odot$ and 
$L_{10} = (L/1.1)/10^{10}h^{-2}L_\odot$.  
If $\alpha(L)$ is independent of $L$, then $M$ determines the 
normalization of the satellite galaxy luminosity function but not 
its shape.  (This is not quite true, because a given $M$ would not 
host satellites more luminous than $L_{\rm min}$, so, strictly 
speaking, it is the faint end shape of the satellite galaxy 
luminosity function which would be independent of halo mass $M$.)  
If $\alpha$ is independent of $L$, then the cumulative function is 
 $\phi_{\rm sat}(\ge L|M)\propto \exp(-\alpha\,L_{10})$ 
at $L_{10}\gg 1$, so the luminosity function itself also
falls as $\exp(-\alpha\,L_{10})$ at  $L_{10}\gg 1$.  
In the other limit, $L_{10}\ll 1$, the cumulative function is 
 $\phi_{\rm sat}(\ge L|M)\propto L_{10}^{-\alpha}$ 
so $\phi_{\rm sat}(L|M)\propto L_{10}^{-\alpha-1}$.  This shows 
explicitly that the satellite galaxy luminosity function should be 
reasonably well fit by a Schechter-like form, even though this form 
played no explicit role in the halo-model parameterization.  Note 
that $L_{10}=1$ is indeed close to the value of $L_*$ associated 
with a Schechter function fit to the SDSS luminosity function 
(Blanton et al. 2003).  

For completeness the luminosity function is 
\begin{eqnarray}
 L\phi_{\rm sat}(L|M) &=& \alpha\,L_{10} \,
   \left[{M_{12}/23\over \exp(L_{10})-1}\right]^{\alpha(L)}\nonumber\\ 
   && \times  
      \left\{{\exp(L_{10})\over \exp(L_{10}) - 1}
             - {\ln(M/M_1)\over L_{10}}
             {{\rm d}\ln\alpha\over {\rm d}\ln L} 
             \right\}
 \label{phisat}
\end{eqnarray}
If $\alpha$ does not depend on $L$, then this simplifies to 
\begin{eqnarray}
 L\phi_{\rm sat}(L|M) &=& \left({M_{12}\over 23}\right)^{\alpha}  
     {\alpha L_{10}\,{\rm e}^{-\alpha L_{10}}\over 
      [1 - \exp(-L_{10})]^{1+\alpha}}.
 \label{phiLsat}
\end{eqnarray}

\section{Galaxies as halo substructure}\label{compareM2L}
Dark matter haloes are expected to have substructure.  
Recent work suggests that numerical simulations now give reliable 
estimates of the subhalo mass function:
\begin{equation}
 {{\rm d}N(m|M)\over {\rm d}m}\, {\rm d}m = 0.01\,
  \left({M\over 10^{12} h^{-1}M_\odot}\right)^{0.1}\,
  \left({M\over m}\right)^{0.9}\,{{\rm d}m\over m}
 \label{dNgao}
\end{equation}
(Gao et al. 2004b).  
Hence, the number of subhaloes more massive than $m$ is  
\begin{eqnarray}
 N(\ge m|M) &=& {0.01\over 0.9}\,
  \left({M\over 10^{12} h^{-1}M_\odot}\right)^{0.1}\,
  \left[\left({M\over m}\right)^{0.9}-1\right]\nonumber\\
  &\approx& {0.01\over 0.9}\,
  \left({M\over 10^{12} h^{-1}M_\odot}\right)^{0.1}\,
  \left({M\over m}\right)^{0.9}.
\end{eqnarray}
If we use $M_1$ to denote the value of $M$ at which the number 
of subhaloes is unity, then the expression above implies that 
\begin{equation}
 \left({M_1\over m}\right) 
  \approx 90\,\left({10^{12} h^{-1}M_\odot}\over m\right)^{0.1} 
  \approx {90\over m_{12}^{0.1}},
\end{equation}
where $m_{12}$ is the subhalo mass in units of $10^{12}h^{-1}M_\odot$.
This shows that the mass required to host at least one subhalo 
of mass $m$ is about 90 times greater than $m$.  
This is substantially larger than the value of $2m$ that one 
might naively have guessed from mass conservation.

Suppose we identify satellite galaxies with subhaloes, and require 
that the relation between satellite galaxy and subhalo mass is 
monotonic and deterministic (e.g. Kravtsov et al. 2004 and 
Conroy et al. 2006, although these authors argue that it is more 
reasonable to use subhalo circular velocity rather than mass).  
Then requiring that 
\begin{equation}
 N(\ge m|M) = \left[{M\over M_1(L)}\right]^{\alpha(L)}
\end{equation}
provides a constraint on the mass-to-light ratio of subhaloes.  
We are particularly interested in quantifying how different this 
ratio is for satellite galaxies compared to centrals.  
We expect a difference simply because the mass required to host 
two galaxies above some luminosity is of order twenty times larger 
that that required to host one, and 20 is much smaller than 90.  

The expression above requires
\begin{equation}
 {M_{12}/90\over m_{12}^{0.9}} 
 = \left[{M_{12}/23\over \exp(L_{10})-1}\right]^{\alpha(L)}
\end{equation}
(recall $L_{10}$ is the luminosity in units of 
$1.1\times 10^{10}\,h^{-2}L_\odot$), which can be rearranged to 
\begin{equation}
 m_{12}^{0.9} =  {M_{12}/90\over (M_{12}/23)^{\alpha(L)}} 
                 \Bigl[\exp(L_{10})-1\Bigr]^{\alpha(L)}.
 \label{satM2L}
\end{equation}
In contrast, the relation for centrals is $M_{12} = \exp(L_{10})-1$ 
(cf. equation~\ref{MLapprox}).  To see what this difference implies, 
it is instructive to consider the case when $\alpha = 1$ independent 
of $L$.  Then 
\begin{eqnarray}
 m_{12} &=& \left({23\over 90}\right)^{10/9}\,
           \Bigl[\exp(L_{10})-1\Bigr]^{10/9}\nonumber\\
        &=& \left[{\exp(L_{10})-1\over \exp(13.643)-1}\right]^{1/9}
           \Bigl[\exp(L_{10})-1\Bigr].
\end{eqnarray}
The first line shows that satellites with $L_{10}=1$ (recall this 
is approximately $L_*$) are 0.233 times less massive than centrals 
of the same luminosity.  The second line shows that satellites are 
less massive than centrals provided $L_{10}<13.643$.  Notice that 
these relations do not depend on the mass $M_{12}$ of the parent 
halo.  If the smaller mass for the satellites is due to tidal 
stripping (this assumes no change in the luminosity of the satellite 
as it falls onto its parent halo), then $\alpha=1$ suggests that, 
on average, a galaxy loses the same fraction of its original dark 
matter halo whether it falls into a small group or a massive cluster.  
If $\alpha\ne 1$, then the mass fraction which survives scales 
with parent halo mass as $M_{12}^{(1-\alpha)/0.9}$, 
although the overall scaling depends on $L_{10}$.  
However, if $\alpha=0.9$, then the mass fraction which survives 
increases with parent halo mass as $M_{12}^{1/9}$, independent of 
luminosity.  
If $\alpha>1$, then the mass fraction which survives 
decreases with increasing $M_{12}$.  
Our neglect of even passive evolution means that we are slightly 
overestimating the masses of the satellites prior to stripping, 
so we are slightly overestimating the mass fraction which is 
stripped.



The analysis above assumes that halo substructure at a given time 
is related to the galaxy population at the same time.  It is not 
obvious that this is reasonable:  equation~(\ref{dNgao}) for the 
subhalo population is calibrated from simulations of dark matter 
clustering.  Some of the subhaloes which disrupt in these simulations 
are expected to survive if the effects of baryons are included.  
This is because baryons cool into the center of their host halo, 
thus inhibiting disruption (Gao et al. 2004a; Diemand et al. 2004; 
Macci\'{o} et al. 2006).  This particularly affects the galaxy 
population close to the halo center:  subhaloes near the host halo 
center tend to be more tidally stripped, making the mass-to-light 
ratio of satellites smallest close to the center, and larger at 
larger radii (Tormen 1997; Hayashi et al. 2003; Gill et al. 2005; 
Gao et al. 2004a; Nagai \& Kravtsov 2005).  
As a consequence of this effect, models which seek to identify 
$z=0$ subhaloes in dark matter only simulations with $z=0$ galaxies 
will have trouble accounting for the `orphan' satellite galaxies 
which should remain after their subhaloes have been disrupted; these 
orphans are expected to contribute to the abundance and small-scale 
clustering of faint galaxies (e.g. Wang et al. 2006).  
Our analysis does not account for this effect, because it will 
only increase the difference between the mass-to-light ratios 
of centrals and satellites of the same luminosity.  

\section{Discussion}
Halo model interpretations of the observed luminosity dependence 
of galaxy clustering suggest that central galaxies in haloes are 
different from all the others---the satellites.  Whereas the 
luminosity of the central object is predicted to be a relatively 
strong function of halo mass (equation~\ref{LMapprox}), the mean 
luminosity of satellite galaxies (i.e., those which are not central 
galaxies) should depend only weakly on halo mass (equation~\ref{LsatM}).  
Since the number of galaxies in a halo is also predicted to increase 
steeply with increasing halo mass, the luminosity of the central galaxy 
is expected to depend strongly on the number of galaxies in a group, 
whereas the average luminosity of the satellites is expected to 
depend little if at all on the group `richness'.  
Figure~\ref{berlindLsat} shows that this prediction is in good 
quantitative agreement with a direct measurement of this trend 
in a catalog of galaxy groups.  

In fact, not just the mean, but the entire shape of the satellite 
galaxy luminosity function, is predicted to be approximately 
independent of halo mass, having a steep power-law form at 
$L\ll 10^{10}h^{-2}L_\odot$, and an exponential cutoff at 
$L\gg 10^{10}h^{-2}L_\odot$ (equations~\ref{cumLsat} 
and~\ref{phiLsat}).  Thus, the satellite luminosity function is 
predicted to have a form like that proposed by Schechter (1976), 
even though this functional form played no role in the halo model 
parametrization.  

The approximate mass independence of $\phi_{\rm sat}(L)$ 
suggests that approaches which use the halo model to infer how 
the number of galaxies above some $L$ depends on halo mass 
(e.g. Zehavi et al. 2005; sometimes called the HOD approach) may 
help simplify halo model decompositions which are based on the 
conditional luminosity function (e.g. Yang et al. 2003; 
Cooray 2006; van den Bosch et al. 2007).  
This is because the CLF approach must separately parametrize 
how the luminosity function of central and satellite galaxies 
depends on halo mass:  the HOD-based analysis here suggests that 
ignoring the mass dependence of the satellite galaxy LF (thus 
reducing the numbers of free parameters to be fitted) should be 
a reasonable first approximation.  

The quantitative parts of our discussion were based on the 
halo model parameters derived by Zehavi et al. (2005) from a 
consideration of clustering as a function of luminosity in the SDSS main galaxy sample, 
and the group catalogs are from Berlind et al. (2006) and Yang et al. (2005a).  
Appendix~A shows that, in some respects, the Zehavi et al. numbers 
are inconsistent with the Berlind et al. catalog.  While this may 
be due to inaccuracies in the group catalog, it may also be due 
to Zehavi et al.'s assumption that $\sigma_8=0.9$ (Berlind et al. and Yang et al.
assumed this same value when calibrating their group-finders).  
While changing $\sigma_8$ within reasonable bounds is unlikely to 
invalidate our findings, we hope that the discussion which follows, 
showing the various implications such measurements can have, will 
motivate a reanalysis of the SDSS main galaxy sample, but with a 
lower value of $\sigma_8$.  

The mass-independence of the satellite luminosity function
has implications for the stellar mass-to-light ratios of satellite galaxies.
Since $M_\ast/L$ is strongly correlated with galaxy color (\textit{e.g.}, Bell et al. 2003),
the approximate independence of $L_\mathrm{sat}$ on halo mass means that 
we should expect that satellite color is an excellent indicator of 
satellite stellar mass, regardless of the mass of the host halo.

The approximate independence of satellite luminosity functions 
on group size also suggests that the extreme value statistics of the 
sort pioneered by Scott (1957) should provide a good description of 
the luminosity function of the most luminous satellites.  
Note that most previous work has used extreme value statistics to 
model the luminosity function of the central rather than satellite 
galaxies (e.g. Bhavsar \& Barrow 1985).  
This is the subject of work in progress (also see Vale \& Ostriker 2007).  
More recent work has phrased this discussion in terms of the luminosity 
gap between the first and second, or second and third ranked galaxies 
in clusters.  For example, Milosavljevic et al. (2006) suggest that the 
gap between first and second 
ranked galaxies is correlated with the dynamical age of the system:  
``fossil'' groups, poor clusters, and rich clusters are distinguished 
by the time since their last major merger, so their luminosity gaps 
differ.  
In addition, van den Bosch et al. (2007) find that the average 
luminosity gap and the fossil group fraction both increase with 
decreasing host halo mass.  

This last finding is particularly easy to understand in the 
context of our results.  At small halo masses, the luminosity of 
the central galaxy grows linearly with halo mass, but the 
growth is only logarithmic at large masses (equation~\ref{LMapprox}). 
On the other hand the number of satellites grows slightly more 
strongly than linearly (equation~\ref{sdssNg}).  If the 
satellite luminosity function is independent of halo mass, 
then massive haloes are allowed more draws from the universal 
satellite luminosity function.  If this function has an 
exponential tail, and equations~(\ref{phisat}) and~(\ref{phiLsat}) 
suggest that it does, then the most luminous of these draws grows 
logarithmically with the number of draws, so it grows logarithmically 
with halo mass.  Thus, the luminosity gap is larger at small masses, 
and decreases at larger masses.  This effect is further helped by 
the fact that 
(i) the satellite luminosity function is not quite independent of 
    halo mass---mean satellite luminosity increases slightly with 
    halo mass; 
(ii) in equation~(\ref{sdssNg}), $M_1(L)\approx 23\,M_{\rm min}(L)$, 
    but the factor of 23 is replaced by a smaller factor at large $L$.  
    In effect, this allows for even more luminous satellites in 
    massive haloes.   

If satellite galaxies are associated with the subhaloes of dark 
matter haloes, then the halo model predicts that central and 
satellite galaxies of the same luminosity should differ in mass by 
factors of about $90/23\sim 4$ (the centrals being more massive), 
whereas the stellar masses at fixed luminosity are unlikely to be 
very different (equation~\ref{satM2L}).  
Weak-lensing analyses should soon be able to test this prediction 
(e.g. Yang et al. 2006), 
as should analyses of satellite dynamics (e.g. McKay et al. 2002).   
In practice, there is likely to be more scatter between subhalo 
mass and luminosity than there is between parent halo mass and 
luminosity; we expect this to alter our conclusions quantitatively 
but not qualitatively.  This too can be checked by lensing analyses.  
At the time or writing, Limousin et al. (2007) have concluded a 
weak-lensing study of five clusters.  They report a difference 
between central and satellite masses (at fixed luminosity) of 
about a factor of 5.
Dynamical mass is proportional to $R\,\sigma^2$, so since
tidal stripping usually does not significantly affect velocity dispersion,
the smaller sizes of cluster galaxies compared to field galaxies
at the same luminosity implies that the satellite galaxies are less massive
by a similar factor.
The measurements are still quite uncertain, but constraints from lensing
are improving. We look forward to more such data, since our 
analysis has shown that such studies are rather closely related to 
studies of the luminosity dependence of galaxy clustering.  

If the difference between the factors of 23 (in the halo model 
description of the mass required to host one satellite) and 90 
(in the subhalo mass function) is associated with mass lost to 
stripping processes as satellites become incorporated into parent 
haloes, then the mass of a satellite prior to stripping is about 
$90/23\sim 4$ times larger than its current mass:  about 80\% of 
its mass is stripped.  This is slightly larger than the $\sim 60\%$
mass-loss factors seen in simulations which only include the 
dark matter component (e.g. Nagai \& Kravtsov 2005).  
Given that the halo-model argument is based on relating the subhalo 
population in simulations to {\em observations}, it is remarkable 
that the two estimates are similar.  

Our analysis of the connection between subhaloes, galaxies and the 
halo model has another interesting consequence.  The mass fraction 
in subhaloes is 
\begin{equation}
 \int_0^M {m\over M}\, {{\rm d}N(m|M)\over {\rm d}m}\, {\rm d}m 
  = 0.1\,\left({M\over 10^{12} h^{-1}M_\odot}\right)^{0.1},
 \label{massfrac}
\end{equation}
where we have used equation~(\ref{dNgao}) for the subhalo mass 
function, $dN/dm$.  If stars only form in sufficiently massive 
objects, the lower limit to this integral may be greater than 
zero:  this will change the quantitative estimates which follow, 
but not the qualitative conclusions.  

Our estimate of the mass lost to stripping processes, when combined 
with equation~(\ref{massfrac}) for the mass fraction in subhaloes, 
leaves about half the mass of a $10^{12}h^{-1}M_\odot$ parent halo 
unaccounted for.  For a $10^{15}h^{-1}M_\odot$ mass halo this 
fraction is about twenty percent.  
For comparison, using the model of subhalo mass-loss due to stripping
in Vale \& Ostriker (2005; see their Figure 1) with equation~(\ref{massfrac}), 
these corresponding fractions are about thirty percent and twenty percent,
when subhaloes with $m>10^{11}h^{-1}M_\odot$ are considered.  (These
mass fraction estimates are highly sensitive to their model of the amount of mass 
stripped from the numerous very low mass haloes, however.)
Presumably this mass is associated with the central galaxy itself, 
and/or with subhaloes that were completely disrupted by the parent 
halo.  If these objects hosted stars, then these stars may have been 
incorporated into the central object, or they may now contribute 
to intracluster light.
Indeed, results from recent studies of intracluster light support 
the idea that much of the light in the ICL comes from the stripping,
disruption, and merging of satellite galaxies (Gonzalez et al. 2005; 
Zibetti et al. 2005).  

Consider a $10^{15}h^{-1}M_\odot$ cluster, for which the halo model 
predicts about eighty percent of the stellar mass is associated with 
satellite galaxies; the rest is in the BCG or in the intracluster medium.  
If the luminosity and color of the BCG are observed (as is the case 
for the SDSS), reasonable assumptions about its stellar mass allow 
one to predict the stellar mass associated with the ICL.
For example, the halo model says the BCG is about 4.5 times more 
luminous than the satellite galaxies brighter than $M_r<-19.9$, 
and that there should be about 70 such satellites.  If the stellar 
mass to light ratio is independent of $L$ (for the red galaxies 
in a cluster this should be a reasonable assumption), then 
the stellar mass of the BCG counts like an additional 4.5 satellites.  
This makes the stellar mass accounted for (74.5/70) times 80\%, 
leaving about 15\% of the stellar mass for the ICL or elsewhere.  

At $10^{14}h^{-1}M_\odot$ the halo model has about 15\% of the 
total mass in subhaloes, whose associated satellite galaxies contain 
about 60\% of the total stellar mass.  There are about 5 satellite 
galaxies brighter than $M_r<-19.9$, and the BCG is a little more 
than 3 times the average luminosity of these satellites.  
This leaves $1 - 0.6 (8/5)$ or about 5\% of the stellar mass for 
the ICL.  Thus, the halo model predicts the ICL fraction to increase 
with host halo mass, in agreement with other recent work 
(Purcell et al. 2007; Murante et al. 2007).
Despite the crudeness of these estimates, we think our discussion 
illustrates how the halo model can be related to recent studies of 
intracluster light. 
This may be particularly useful in view of current uncertainties 
about the fate of `orphan' satellites in simulations 
(Gao et al. 2004a; Conroy et al. 2007).

\section*{Acknowledgements}
We thank Andreas Berlind for helpful discussions about his 
group catalog, and Frank van den Bosch for providing the group 
catalog of Yang et al., and for discussions about our conclusions.  
RAS thanks Sebastian Jester for help with using SDSS CasJobs
to obtain photometric information for groups with fiber collided galaxies.
This work was supported in part by NSF grant 0520647.

Funding for the SDSS and SDSS-II has been provided by the Alfred P. Sloan Foundation, the Participating Institutions, the National Science Foundation, the U.S. Department of Energy, the National Aeronautics and Space Administration, the Japanese Monbukagakusho, the Max Planck Society, and the Higher Education Funding Council for England. The SDSS Web Site is http://www.sdss.org/.

The SDSS is managed by the Astrophysical Research Consortium for the Participating Institutions. The Participating Institutions are the American Museum of Natural History, Astrophysical Institute Potsdam, University of Basel, University of Cambridge, Case Western Reserve University, University of Chicago, Drexel University, Fermilab, the Institute for Advanced Study, the Japan Participation Group, Johns Hopkins University, the Joint Institute for Nuclear Astrophysics, the Kavli Institute for Particle Astrophysics and Cosmology, the Korean Scientist Group, the Chinese Academy of Sciences (LAMOST), Los Alamos National Laboratory, the Max-Planck-Institute for Astronomy (MPIA), the Max-Planck-Institute for Astrophysics (MPA), New Mexico State University, Ohio State University, University of Pittsburgh, University of Portsmouth, Princeton University, the United States Naval Observatory, and the University of Washington.

\appendix

\renewcommand{\thefigure}{\Alph{appfig}\arabic{figure}}
\setcounter{appfig}{1}
\begin{figure}
 \centering
 \includegraphics[width=1.7\hsize]{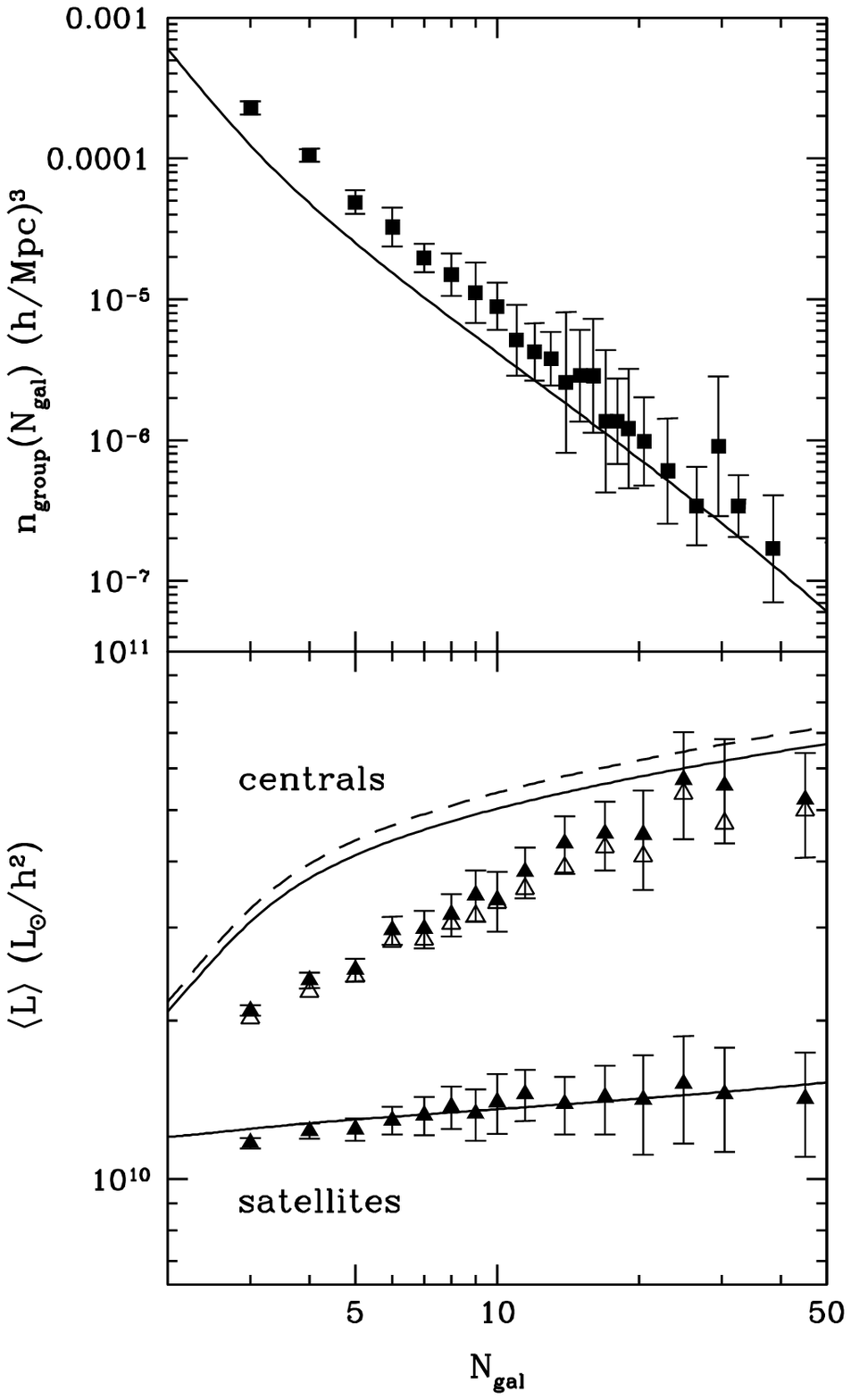} 
 \caption{The halo model of Zehavi et al. (2005) underestimates 
          the abundances of $M_r\le -19.9$ groups (top), 
          and overestimates the typical luminosities of central galaxies (bottom).
          For $\langle L_{\rm cen}|N_{\rm gal}\rangle$, the open points and dashed curve
          show the original measurement and halo model prediction, and the solid points
          and curve show the fiber collision-corrected ones, respectively (see text for details).
          Whether or not the discrepancies in the multiplicity function and
          central galaxy luminosities are a function of
          assuming $\sigma_8=0.9$ is the subject of work in progress.  }
 \label{problems}
\end{figure}


\section{Inconsistencies between the halo model of Zehavi et al. 
and the group catalog of Berlind et al.}
The main text used the halo model parameters of Zehavi et al. (2005).  
Figure~\ref{problems} shows that this halo model underpredicts the 
abundance of groups in Berlind et al.'s $M_r\le -19.9$ group catalog, 
and it overpredicts the central galaxy 
luminosities in these groups.
That is, a model that correctly describes the abundance and clustering
of SDSS galaxies as a function of luminosity is not consistent with the
multiplicity function and central luminosities of this catalog of SDSS galaxy groups.
We have performed the same analysis with Berlind et al.'s $M_r\le -19$ 
group catalog and have found the same discrepancies.

Berlind et al. (2006) have used mock catalogs in order to statistically
correct for fiber collisions, where the thickness of the spectroscopic fibers 
means that galaxies closer than 55" on the sky will be missing spectra and redshifts.
Since such close galaxies tend to be located in dense regions, fiber collisions could
still potentially be the source of the discrepancy of the central galaxy luminosities.
Fiber collided galaxies have been 
given the redshifts and magnitudes of their nearest neighbors, so central galaxy
luminosities may be underestimates.  To test this effect, we obtained the SDSS
photometric data for the groups with fiber collided galaxies in the Berlind et al.
group catalog used in Figure~\ref{berlindLsat}.  We k-corrected, evolution-corrected,
and extinction-corrected them the same way as was done with the galaxies with spectra.
We repeated our analysis of $\langle L_{\rm cen}|N_{\rm gal}\rangle$,
and found it was indeed biased low, but only by 7\% on average.
In addition, given knowledge of the frequency of fiber collisions
as a function of $N_{\rm gal}$, we estimated the effect by modifying 
equation (\ref{LcenN}), and we similarly found a bias of about 7\%.
The fiber collision-corrected measurement and halo model prediction
of $\langle L_{\rm cen}|N_{\rm gal}\rangle$ are shown in the lower
panel of Figure~\ref{problems}, and it is clear that the effect of
fiber collisions is smaller than the uncertainties in the measurements
and less significant than the discrepancy with the model.

We believe the discrepancies in the multiplicity function and central galaxy luminosities
in the figure may be a consequence of Zehavi et al.'s assumption 
that $\sigma_8=0.9$, which is somewhat larger than the value 
$\sigma_8\approx 0.8$ suggested by more recent analyses of other 
data sets (e.g. van den Bosch et al. 2007).  
We argue below that $\sigma_8=0.8$ is likely to reduce the 
discrepancies shown in Figure~\ref{problems} without changing the 
main point of our paper: that satellite luminosities depend only 
weakly on halo mass, and that the dark haloes which surround central 
galaxies are more massive than those which surround satellites of 
the same luminosity.  

If $\sigma_8$ is lower, then the number of massive haloes is lower 
as is the clustering of the dark matter.  
To match the observed abundances and clustering in the main galaxy 
catalog, the number of galaxies in more massive haloes must be increased 
relative to the numbers from Zehavi et al.  This will have the effect 
of increasing the number of groups with large $N$, thus reducing the 
discrepancy seen in the top panel of Figure~\ref{problems}.  
In addition, if $\sigma_8$ is smaller, then a given value of $N$ would 
be associated with smaller mass haloes, making the mass to light 
ratio of haloes smaller.  Such a change would help alleviate a 
source of tension between halo models and observations:  
the mass-to-light ratio of $10^{14}h^{-1}M_\odot$ haloes in our model 
with the Zehavi et al. halo occupation distribution is
about $900\pm100$ for $M_r<-20$, 
and although this is
consistent with similar models for the same cosmology (Tinker et al. 2005),
recent observations have lower mass-to-light ratios that require lower values
of $\sigma_8$ and $\Omega_M$ to model them consistently with the abundance
and clustering of galaxies (van den Bosch et al. 2007).

Also, Berlind et al. (2007) have argued that the clustering of these 
groups are indeed better fit by a cosmological model in which 
$\sigma_8=0.8$, though they do not show if the associated group 
abundances are consistent with this value (Sheth \& Tormen 1999 show 
that, in hierarchical models, abundance and clustering are very 
closely related, so it is likely that the abundances are also 
better fit by the lower $\sigma_8$ value).  

For all these reasons, it is likely that repeating Zehavi et al.'s 
(2005) analysis of the SDSS main galaxy sample, 
but with a lower value of $\sigma_8$, would be very interesting.  
While such a reanalysis is beyond the scope of this work, we note 
that our main results are unlikely to be altered by such a change.  
This is because changing $\sigma_8$ by $\sim 10\%$ is unlikely to 
change the fact that $\alpha$ in equation~(\ref{sdssNg}) is only a 
weak function of $L$, and it is this weak dependence which makes 
satellite luminosities only weakly dependent on halo mass 
(c.f. discussion following equation~\ref{LsatM}).  

\addtocounter{figure}{1}
\begin{figure}
 \centering
 \includegraphics[width=0.9\hsize]{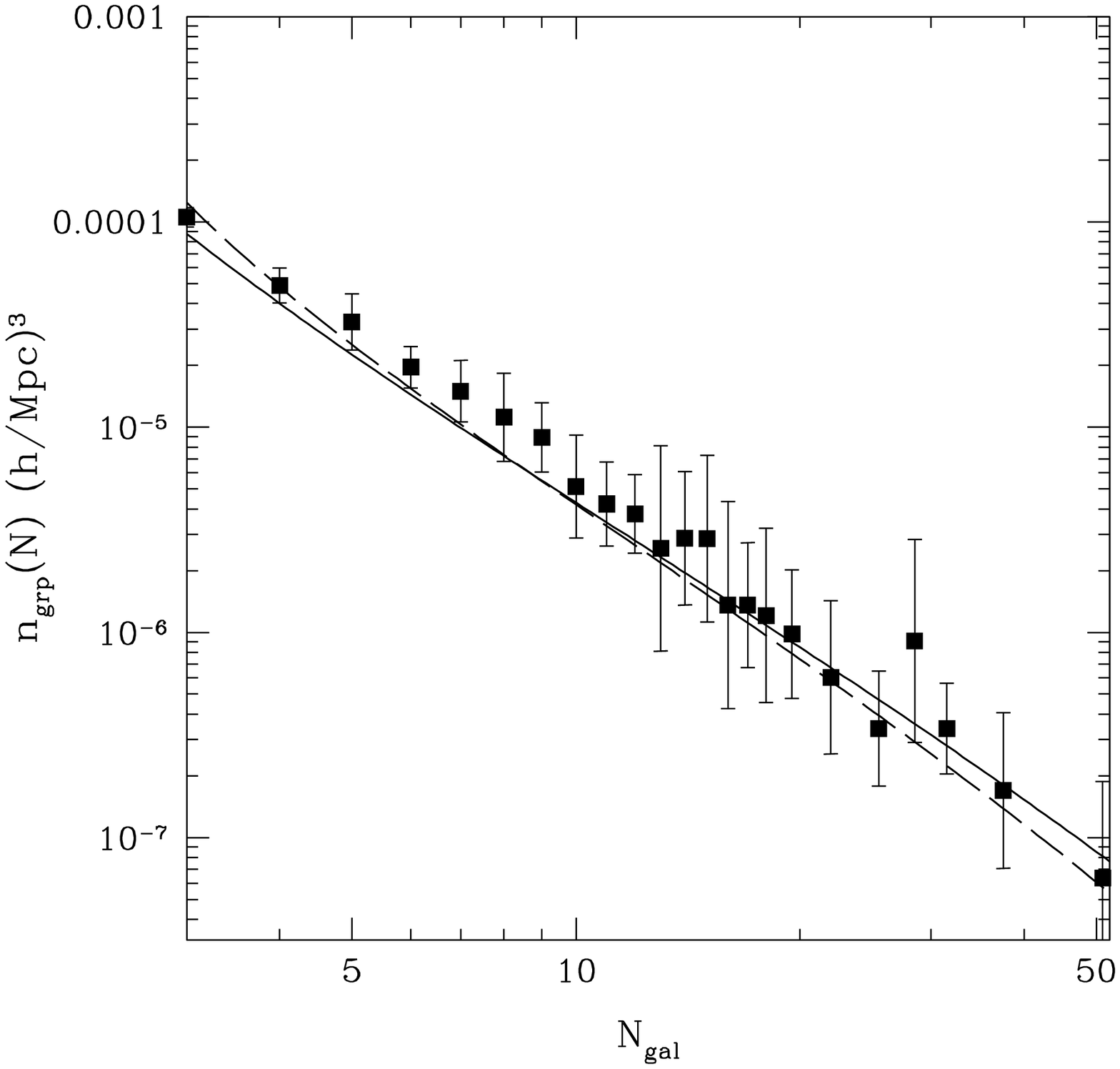} 
 \caption{Same as Figure~\ref{problems} but with $N_\mathrm{gal}\rightarrow(N_\mathrm{gal}-1)$
          for the group catalog measurement.
          Now there is good agreement with the halo model (dashed curve).
          For comparison, the halo model prediction for $\sigma_8=0.8$ and $\Omega_M=0.25$,
          constrained by the correlation function and luminosity function at
          $\Omega_M=0.3$, is also shown (solid curve). 
         }
 \label{FoFproblems}
\end{figure}

Finally, it is important to note that the friends-of-friends (FoF) algorithm adopted by
Berlind et al. (2006) is not an unbiased group-finder, especially at low $N_\mathrm{gal}$.
As they show in their Figures 12 (in real space) and 14 (in redshift space) in their
appendix, for a particular choice of linking length, the FoF algorithm correctly recovers 
the ``true'' multiplicity function in mock catalogs only for $N_\mathrm{gal}>10$.
At $N_\mathrm{gal}\leq10$, the algorithm is increasingly biased with decreasing 
$N_\mathrm{gal}$: it finds too many groups, merging too many lower $N_\mathrm{gal}$
groups and isolated galaxies together.  The problem is worse in redshift space, where 
$N_\mathrm{gal}<8$ groups are over-counted by $50\%$ or more, which clearly shows that the 
small error bars at low $N_\mathrm{gal}$ in Figure~\ref{problems} are misleading.
In Figure~\ref{FoFproblems} we show that if, on average, the FoF group-finder tends to 
link one galaxy too many to each group, then the resultant multiplicity function
agrees very well with the halo model prediction (solid curve).
For comparison, we also show the halo model prediction for $\sigma_8=0.8$ and 
$\Omega_M=0.25$ (dashed curve), with the halo occupation distribution constrained by the
correlation function and luminosity function at $\Omega_M=0.3$.
In particular, for this calculation we are assuming that the change to $\Omega_M$ is
sufficiently small that the $\phi(L)$ (Blanton et al. 2003) and $w_p(r_p|L)$ (Zehavi et al. 2005)
to which the halo model is fit are unchanged.
This is not an unreasonable assumption, since the luminosity functions and correlation functions 
of volume-limited SDSS catalogs have been found to be not significantly affected by such 
a small change in cosmology (F. C. van den Bosch, private communication). 
We then changed $\sigma_8$ in the halo model, thus changing the linear power spectrum and 
the halo mass function, and found what new halo occupation distribution is required that 
still correctly describes $\phi(L)$ and $w_p(r_p|L)$.
The resulting best-fitting halo occupation distributions for $M_r<-20$ have somewhat 
similar $M_\mathrm{min}(L)$, but with significantly lower $M_1(L)$ and a steeper 
$\langle N_\mathrm{sat}|M\rangle$ slope (see equation~\ref{sdssNg}).
However, the resulting multiplicity function predicted by the halo model is similar for both
$\sigma_8=0.8$ and $0.9$, which suggests that the discrepancy in Figure~\ref{problems}
may be due as much to the group-finder itself as to the assumed cosmology.

\label{lastpage}
\end{document}